\documentclass[review]{elsarticle}

\pdfoutput=1

\usepackage{amsmath}									
\usepackage{amsbsy}
\usepackage{color}
\usepackage{esint}
\usepackage{comment}
\usepackage{graphicx}

\usepackage[T1]{fontenc}
\usepackage[bitstream-charter]{mathdesign}
\usepackage[latin1]{inputenc}							

\usepackage{caption}
\usepackage{enumitem}

\usepackage{lineno}
\modulolinenumbers[5]

\newcommand*{\E}{\mathbb{E}}

\renewcommand*{\doteq}{:=}

\newcommand*{\Var}{\operatorname{Var}}

\newcommand*{\EE}{\mathbb E}
\newcommand*{\PP}{\mathbb P}

\newcommand*{\bbN}{\mathbb N}

\newcommand*{\bbR}{\mathbb R}

\newcommand*{\cI}{\mathcal I}

\newcommand*{\cU}{\mathcal U}

\DeclareMathOperator{\sgn}{sgn}

\newtheorem{theorem}{Theorem}
\newtheorem{lemma}{Lemma}

\newtheorem{remark}{Remark}
\newtheorem{proposition}{Proposition}
\newproof{proof}{Proof}

\journal{Mathematics and Computers in Simulation}

\begin{document}
\begin{frontmatter}
\title{Stratified regression-based variance reduction approach for weak approximation schemes}

 \author[due,iitp]{D.~Belomestny\corref{cor1}\fnref{fn1}}
 \ead{denis.belomestny@uni-due.de}
 
\author[pwc]{S.~H\"afner}
\ead{stefan.haefner@de.pwc.com}
 
 
 \author[due]{M.~Urusov}
 \ead{mikhail.urusov@uni-due.de}
 
 \cortext[cor1]{Corresponding author}
\fntext[fn1]{The research by Denis Belomestny was made in IITP RAS and supported by Russian Scientific Foundation grant (project N 14-50-00150).}
 
 \address[due]{Duisburg-Essen University, Essen, Germany}
\address[iitp]{IITP RAS, Moscow, Russia}
\address[pwc]{PricewaterhouseCoopers GmbH, Frankfurt, Germany}

\begin{abstract}
In this paper we suggest a modification of the regression-based variance reduction approach recently proposed in Belomestny et al~\cite{belomestny2016variance}. This modification is based on the stratification  technique and allows for  a further significant variance reduction. 
The performance  of the proposed approach  is  illustrated by several numerical \mbox{examples}.
\end{abstract}

\end{frontmatter}


\pagestyle{myheadings}
\thispagestyle{plain}
\markboth{D.~Belomestny, S.~H\"afner, M.~Urusov}{Stratified regression-based variance reduction approach}

\section{Introduction}
Let \(T>0\) be a fixed time horizon.
Consider a $d$-dimensional diffusion process
$(X_t)_{t\in[0,T]}$
defined by the It\^o stochastic differential equation
\begin{align}\label{x_sde}
dX_t
=\mu(X_t)\,dt
+\sigma(X_t)\,dW_{t},
\quad X_{0}=x_0\in\mathbb{R}^d,
\end{align}
for Lipschitz continuous functions
\(\mu\colon\mathbb{R}^d\to\mathbb{R}^d\)
and
\(\sigma\colon\mathbb{R}^d
\to\mathbb{R}^{d\times m}\),
where \((W_t)_{t\in[0,T]}\)
is a standard \(m\)-dimensional Brownian motion.
Suppose we want to compute the expectation 
\begin{align}\label{eq:2512a0}
u(t,x)\doteq\E[ f(X_{T}^{t,x})],
\end{align}
where $X^{t,x}$ denotes the solution to~\eqref{x_sde} started
at time $t$ in point~$x$. The standard Monte Carlo (SMC)
approach for computing \(u(0,x)\) at a fixed point \(x\in \bbR^d\) basically consists of three steps. First, an approximation \(\overline{X}_T\) for \(X^{0,x}_T\) is constructed via a time discretisation in equation~\eqref{x_sde} (we refer to~\cite{KP} for a nice overview of various discretisation schemes). Next, \(N_0\) independent copies of the approximation \(\overline{X}_T\)  are generated, and, finally, a Monte Carlo estimate \(V_{N_0}\) is defined as the average of the values of \(f\) at simulated points:
\begin{align}\label{eq:2512a1}
V_{N_0}\doteq\frac{1}{N_0}\sum_{i=1}^{N_0} f\Bigl(\overline{X}_T^{(i)}\Bigr).
\end{align} 
In the computation of $u(0,x)=\EE [f(X^{0,x}_T)]$
by the SMC approach
there are two types of error inherent:
the discretisation error
$\EE [f(X^{0,x}_T)]-\EE [f(\overline X_{T})]$
and the Monte Carlo (statistical) error,
which results from the substitution of
$\EE [f(\overline{X}_{T})]$
with the sample average~$V_{N_0}$.
The aim of variance reduction methods is to reduce the statistical error.  For example, in the so-called control variate variance reduction approach
one looks for a random variable
\(\xi\) with \(\EE \xi=0\),
which can be simulated,
such that
the variance of the difference
\(f(\overline{X}_{T})-\xi\) is minimised, that~is,
\begin{align*}
\Var[f(\overline{X}_{T})-\xi]\to\min\text{ under } \EE\xi=0.
\end{align*}
Then one uses the sample average
\begin{align}\label{eq:2512a2}
V_{N_0}^{CV}\doteq\frac{1}{N_0}\sum_{i=1}^{N_0}
\left[f\Bigl(\overline{X}_T^{(i)}\Bigr)-\xi^{(i)}\right]
\end{align}
instead of~\eqref{eq:2512a1} to approximate $\EE [f(\overline{X}_{T})]$.
The use of control variates for computing expectations of functionals of diffusion processes via Monte Carlo  was initiated by Newton~\cite{newton1994variance} and further developed in Milstein and Tretyakov~\cite{milstein2009practical}. Heath and Platen~\cite{heath2002variance} use the integral representation to construct unbiased variance-reduced estimators.
In Belomestny et al~\cite{belomestny2016variance} a novel regression-based approach for the construction of control variates, which reduces the variance of  the approximated functional \(f(\overline{X}_{T})\) was proposed.
As shown in~\cite{belomestny2016variance},
the ``Monte Carlo approach with
the Regression-based Control Variate''
(abbreviated below as ``RCV approach'')
as well as its enhancement,
called ``recursive RCV (RRCV) approach'',
are able to achieve a higher order convergence
of  the resulting variance to zero,
which in turn leads to a significant complexity reduction
as compared to the SMC algorithm.
The RCV approaches become especially simple
in the case of the so-called weak approximation schemes,
i.e.,\ the schemes, where simple random variables
are used in place of Brownian increments,
and which became quite popular in recent years.
In this paper we further enhance the performance
of the RRCV algorithm by combining it with stratification.
The idea of the resulting \emph{stratified RCV (SRCV)}
algorithm is based on partitioning of the state space
into a collection of sets \(\mathcal{A}_1,\ldots,\mathcal{A}_p\)
and then performing conditional regressions
separately on each set. It turns out that
by choosing \(\mathcal{A}_1,\ldots,\mathcal{A}_p\)
to be the level sets of the discrete-valued random variables
used in the weak approximation scheme,
we can achieve a further variance reduction effect
as compared to the original approach
in~\cite{belomestny2016variance}. 
The paper is organised as follows. 
In Section~\ref{sec:2}, the SRCV algorithm is introduced
and compared with the RCV and RRCV ones.
The complexity analysis of the SRCV algorithm
is conducted in Section~\ref{sec:3}.
Section~\ref{sec:4} is devoted to the simulation study.
Necessary proofs are collected in Section~\ref{sec:proofs}.

\section{SRCV approach and its differences with RCV and RRCV ones}
\label{sec:2}
In what follows $J\in\bbN$
denotes the time discretisation parameter.
We set $\Delta\doteq T/J$
and consider discretisation schemes
denoted by $(X_{\Delta,j\Delta})_{j=0,\ldots,J}$,
which are defined on the grid $\{j\Delta:j=0,\ldots,J\}$.
In Sections~\ref{sec:21} and~\ref{sec:22}
we consider weak schemes of order~$1$.
In this setting we recall the RCV and RRCV algorithms,
introduce the SRCV algorithm and explain
how it compares to the RCV and RRCV ones.
In Section~\ref{sec:23}
we briefly discuss the case of weak schemes of order~$2$.

\subsection{RCV algorithm for first order schemes}
\label{sec:21}
Let us consider a weak scheme of order~$1$,
where $d$-dimensional approximations
$X_{\Delta,j\Delta}$,
$j=0,\ldots,J$, satisfy $X_{\Delta,0}=x_0$ and
\begin{align}
\label{eq:scheme_structure_md}
X_{\Delta,j\Delta}=
\Phi_{\Delta}(X_{\Delta,(j-1)\Delta},\xi_j),
\quad j=1,\ldots,J,
\end{align}
for some functions
$\Phi_{\Delta}\colon\bbR^{d+m}\to\bbR^d$,
with $\xi_j=(\xi_j^1,\ldots,\xi_j^m)$, $j=1,\ldots,J$,
being $m$-dimensional
i.i.d.\ random vectors with i.i.d.\ coordinates
satisfying
\begin{align*}
\PP\left(\xi_j^k=\pm1\right)=\frac12, \quad k=1,\ldots,m.
\end{align*}
An important particular case is the weak Euler scheme (also called the
\emph{simplified weak Euler scheme}
in \cite[Section~14.1]{KP}), which is given by
\begin{align}\label{eq:PhiK=1}
\Phi_{\Delta}(x,y)
=x+\mu(x)\,\Delta+\sigma(x)\,y\,\sqrt{\Delta}.
\end{align}
The RCV approach of~\cite{belomestny2016variance}
essentially relies on the following representation,
which has some resemblance to the
discrete-time Clark-Ocone formula
(see e.g.~\cite{privault2002discrete}).

\begin{theorem}
\label{th:weak_md01}
It holds
\begin{align}
\label{eq:repr02}
f(X_{\Delta,T})=\EE f(X_{\Delta,T})+
\sum_{j=1}^J
\sum_{k\in\{0,1\}^m\setminus\{0\}}
a_{j,k}(X_{\Delta,(j-1)\Delta})
\prod_{i=1}^m (\xi_j^i)^{k_i},
\end{align}
where $k=(k_1,\ldots,k_m)$ and
$0=(0,\ldots,0)$ (in the second summation).
Moreover, the coefficients
$a_{j,k}\colon\bbR^d\to\bbR$
can be computed by the formula
\begin{align}
\label{eq:coef05}
a_{j,k}(x)
=\EE\left[\left.
f(X_{\Delta,T}) \prod_{i=1}^m (\xi_j^i)^{k_i} 
\,\right|\,
X_{\Delta,(j-1)\Delta}=x
\right]
\end{align}
for all $j$ and $k$ as in~\eqref{eq:repr02}.
\end{theorem}

Theorem~\ref{th:weak_md01}
is an equivalent reformulation
of Theorem~3.1 in~\cite{belomestny2016variance}.

\paragraph{Discussion}
Under appropriate conditions on the functions
$f$, $\mu$ and~$\sigma$
(see e.g.\ Theorem~2.1 in~\cite{MilsteinTretyakov:2004})
the discretisation error
$\EE[f(X_T)]-\EE[f(X_{\Delta,T})]$
for the scheme~\eqref{eq:PhiK=1}
is of order~$\Delta$ (first order scheme).
Furthermore, by Theorem~\ref{th:weak_md01}, with
\begin{align}
\label{eq:2909a2}
M^{(1)}_{\Delta,T}\doteq
\sum_{j=1}^J
\sum_{k\in\{0,1\}^m\setminus\{0\}}
a_{j,k}(X_{\Delta,(j-1)\Delta})
\prod_{i=1}^m (\xi_j^i)^{k_i},
\end{align}
we have $\EE\left[M^{(1)}_{\Delta,T}\right]=0$
and $\Var\left[f(X_{\Delta,T}) - M^{(1)}_{\Delta,T}\right]=0$,
that is, $M^{(1)}_{\Delta,T}$
is a perfect control variate
reducing the statistical error down to zero.
However, in practice we cannot
simulate $M^{(1)}_{\Delta,T}$
because the coefficients $a_{j,k}$
are generally unknown.
In the RCV algorithm, we construct a
practically implementable control variate
$\tilde{M}^{(1)}_{\Delta,T}$
of the form~\eqref{eq:2909a2}
with regression-based estimates
$\tilde{a}_{j,k}\colon\bbR^d\to\bbR$
of the functions $a_{j,k}$.
It is worth noting that the sample average of
$f(X_{\Delta,T}^{(i)})-\tilde M_{\Delta,T}^{(1),(i)}$
(cf.~\eqref{eq:2512a2})
should be computed on the paths
independent of those used to construct $\tilde a_{j,k}$.
This ensures that $\EE\bigl[\tilde{M}^{(1)}_{\Delta,T}\bigr]=0$,
and, thus, that $\tilde M_{\Delta,T}^{(1)}$ is a valid control variate
(because of the martingale transform structure
in~\eqref{eq:2909a2}).

\subsection{RRCV and SRCV algorithms for first order schemes}
\label{sec:22}
The coefficients given by~\eqref{eq:coef05}
can be approximated using various regression algorithms.
From a computational point of view it is however advantageous to look for other representations,
which only involve regressions over one time step
(notice that in~\eqref{eq:coef05}
the regression is performed over \(J-j+1\) time steps).
To this end, for $j\in\{1,\ldots,J\}$, we introduce the functions
\begin{align}
\label{tower:q}
q_j(x)\doteq\EE[f(X_{\Delta, T})|X_{\Delta,j\Delta}=x].
\end{align}
The next result is Proposition~3.3
of~\cite{belomestny2016variance}.

\begin{proposition}
\label{prop:2202a1}
We have $q_J\equiv f$ and, for each $j\in\{2,\ldots,J\}$,
\begin{align}
\label{eq:2408a1}
q_{j-1}(x)=&\EE\bigl[q_j(X_{\Delta, j\Delta})|X_{\Delta,(j-1)\Delta}=x\bigr]=\frac{1}{2^m}
\sum_{y\in\{-1,1\}^m} q_j(\Phi_\Delta(x,y)).
\end{align}
Moreover,
for all $j\in\{1,\ldots,J\}$ and $k=(k_i)\in\{0,1\}^m\setminus\{0\}$
(with $0\equiv(0,\ldots,0)$),
the functions $a_{j,k}(x)$ in~\eqref{eq:coef05}
can be expressed in terms of the functions
$q_j(x)$ as follows:
\begin{align}
\label{eq:coef05a}
a_{j,k}(x)=\frac{1}{2^m}
\sum_{y=(y_1,\ldots,y_m)\in\{-1,1\}^m}
\left[\prod_{i=1}^m y_i^{k_i}\right] q_j(\Phi_\Delta(x,y)).
\end{align}
\end{proposition}

The first equality in~\eqref{eq:2408a1} shows
that we can recursively approximate the functions $q_j(x)$
via regressions over one time step only.
This gives the RRCV algorithm
of~\cite{belomestny2016variance}:
first compute regression-based approximations
$\tilde q_j(x)$ of the functions $q_j(x)$
(via regressions over one time step
based on the first equality in~\eqref{eq:2408a1}),
then obtain approximations
$\tilde a_{j,k}(x)$
of the functions $a_{j,k}(x)$
via~\eqref{eq:coef05a}
with $q_j$ being replaced by~$\tilde q_j$,
and, finally, construct the control variate $\tilde{M}^{(1)}_{\Delta,T}$
using~\eqref{eq:2909a2}
with $a_{j,k}(x)$
being replaced by~$\tilde a_{j,k}(x)$.

To introduce the SRCV algorithm, we first define
functions $h_{j,y}$, for all $j\in\{1,\ldots,J\}$
and $y\in\{-1,1\}^m$, by the formula
\begin{align}
\label{eq_h}
h_{j,y}(x)\doteq q_j(\Phi_\Delta(x,y))
=\EE[q_j(X_{\Delta,j\Delta})|X_{\Delta,(j-1)\Delta}=x,\xi_j=y]
\end{align}
(the second equality is straightforward)
and observe that the knowledge of these functions
for some $j$ and all $y$ provides us with
the functions $q_{j-1}$ and $a_{j,k}$,
$k\in\{0,1\}^m\setminus\{0\}$,
via the second equality in~\eqref{eq:2408a1}
and via~\eqref{eq:coef05a}.
Inspired by this observation
together with the second equality in~\eqref{eq_h},
we arrive at the idea of the stratified regression:
approximate each function
$h_{j,y}(x)$ via its projection on a given set
of basis functions $\psi_1(x),\ldots,\psi_K(x)$.
In detail, the SRCV algorithm consists of two phases:
``training phase'' and ``testing phase''.

\emph{Training phase of the SRCV algorithm:}
First, simulate a sufficient number $N$ of
(independent) ``training paths'' of the discretised diffusion.
Let us denote the set of these $N$ paths by~$D_N^{tr}$:
\begin{align}\label{eq:2712a1}
D_N^{tr}\doteq\left\{
(X_{\Delta,j\Delta}^{tr,(i)})_{j=0,\ldots,J}:
i=1,\ldots,N\right\}
\end{align}
(the superscript ``tr'' comes from ``training'').
Next, proceed as follows.

\emph{Step~1.}
Set $j=J$, $\tilde q_j=f$.
Compute the values
$\tilde q_j(X_{\Delta,j\Delta}^{tr,(i)})$
on all training paths ($i=1,\ldots,N$).

\emph{Step~2.}\label{step2}
For all $y\in\{-1,1\}^m$, construct regression-based
approximations $\tilde h_{j,y}$ of the functions~$h_{j,y}$
(via regressions over one time step
based on the second equality in~\eqref{eq_h}
with $q_j$ being replaced by~$\tilde q_j$).
In fact, only training paths with $\xi_j=y$
are used to construct~$\tilde h_{j,y}$.

\emph{Step~3.}
Using the approximations $\tilde h_{j,y}$
for all $y\in\{-1,1\}^m$, via~\eqref{eq:coef05a}
compute the coefficients $\alpha_1,\ldots,\alpha_K$
in the representations $\alpha_1\psi_1+\ldots+\alpha_K\psi_K$
for the approximations $\tilde a_{j,k}$,
$k\in\{0,1\}^m\setminus\{0\}$.
Note  that the cost of computing any of $\tilde a_{j,k}(x)$
at any point $x$ will be of order~$K$.
Furthermore, again using $\tilde h_{j,y}$
for all $y\in\{-1,1\}^m$,
compute the values
$\tilde q_{j-1}(X_{\Delta,(j-1)\Delta}^{tr,(i)})$
on all training paths ($i=1,\ldots,N$)
via the second equality in~\eqref{eq:2408a1}.

\emph{Step~4.}
If $j>1$, set $j=j-1$ and go to step~2.

Thus, after the training phase is completed,
we have the approximations
$\tilde a_{j,k}(x)$ of $a_{j,k}(x)$
for all $j\in\{1,\ldots,J\}$ and $k\in\{0,1\}^m\setminus\{0\}$.
Let us emphasise that, in fact,
\begin{align}\label{eq:2712a2}
\tilde a_{j,k}(x)=\tilde a_{j,k}(x,D_N^{tr}),
\end{align}
that is, our approximations are random
and depend on the simulated training paths.

\emph{Testing phase of the SRCV algorithm:}
Simulate $N_0$ ``testing paths''
$(X_{\Delta,j\Delta}^{(i)})_{j=0,\ldots,J}$,
$i=1,\ldots,N_0$,
that are independent from each other
and from the training paths and construct
the Monte Carlo estimate
\begin{align}\label{eq:2812b0}
\frac{1}{N_0}\sum_{i=1}^{N_0}
\left[f\Bigl(X_{\Delta,T}^{(i)}\Bigr)-\tilde M_{\Delta,T}^{(1),(i)}\right]
\end{align}
(cf.~\eqref{eq:2512a2}), where
$\tilde M_{\Delta,T}^{(1),(i)}$
is given by
\begin{align}\label{eq:2812b1}
\tilde M^{(1),(i)}_{\Delta,T}\doteq
\sum_{j=1}^J
\sum_{k\in\{0,1\}^m\setminus\{0\}}
\tilde a_{j,k}(X_{\Delta,(j-1)\Delta}^{(i)},D_N^{tr})
\prod_{l=1}^m (\xi_j^{l,(i)})^{k_l}
\end{align}
(cf.~\eqref{eq:2909a2}).

\paragraph{Discussion}
Let us briefly discuss the main differences
between the RRCV and SRCV algorithms.
In the training phase of the RRCV algorithm
the functions $q_j$, $j\in\{1,\ldots,J\}$,
are approximated recursively via regressions
using the first equality in~\eqref{eq:2408a1}
(the second equality in~\eqref{eq:2408a1}
is not used at all), and the approximations
are linear combinations of $K$ basis functions
$\psi_1,\ldots,\psi_K$.
This allows to get the control variate
in the testing phase via the formula like~\eqref{eq:2812b1}
with the coefficients $\tilde a_{j,k}$ constructed
\emph{on the testing paths} via~\eqref{eq:coef05a}
with approximated in the training phase functions~$q_j$.
On the contrary, in the training phase of the SRCV
algorithm regressions are based on
the second equality in~\eqref{eq_h},
and we get approximations for all functions
$h_{j,y}$ ($\equiv q_j(\Phi_\Delta(\cdot,y))$),
$j\in\{1,\ldots,J\}$, $y\in\{-1,1\}^m$,
where the approximations $\tilde h_{j,y}$
are again linear combinations of $K$ basis functions
$\psi_1,\ldots,\psi_K$
(notice that what we now need from~\eqref{eq:2408a1}
is the second equality but not the first one).
Having the approximations $\tilde h_{j,y}$,
we get the approximations of the functions
$\tilde a_{j,k}$ via~\eqref{eq:coef05a}
as linear combinations of $\psi_1,\ldots,\psi_K$
already in the training phase, while the testing phase
is completely described by
\eqref{eq:2812b0}--\eqref{eq:2812b1}.
Let us compare the computational costs
of the RRCV and SRCV algorithms. 
For the sake of  simplicity we restrict our attention to the case of  ``large''
parameters\footnote{\label{ft:3012a1}We need to have
$J\to\infty$, $K\to\infty$, $N\to\infty$, $N_0\to\infty$
in order to make both the discretisation and the
statistical error tend to zero (see Section~\ref{sec:3}
for more detail).}
$J$, $K$, $N$ and $N_0$
as well as at the ``big'' constant\footnote{In contrast to
$J$, $K$, $N$ and $N_0$, the value $c_m\doteq2^m$
is fixed, but can be relatively big (compared to other
involved constants such as e.g.\ $d$ or~$m$).
Notice that $c_m$ is the number of scenarios
that the random variables $\xi_j$ can take,
and it comes into play via formulas
like~\eqref{eq:2812b1} ($J(c_m-1)$ summands)
or~\eqref{eq:coef05a} ($c_m$ summands).}
$c_m\doteq2^m$
ignoring other constants such as e.g.\ $d$ or~$m$.
As for the RRCV algorithm,
$J$ regressions with $N$ training paths
and $K$ basis functions result in
the cost of order $JK^2N$,
while the cost of the testing phase is of
order\footnote{\label{ft:3013a1}Naive implementation
of the testing phase in the RRCV algorithm
via~\eqref{eq:coef05a} and~\eqref{eq:2909a2}
gives the cost order $JKc_m(c_m-1)N_0$.
To get $JKc_mN_0$,
one should implement~\eqref{eq:coef05a}
on the testing paths in two steps:
first, for all
$i\in\{1,\ldots,N_0\}$,
$j\in\{1,\ldots,J\}$
and $y\in\{-1,1\}^m$,
compute the values
$\tilde q_j(\Phi_\Delta(X_{\Delta,(j-1)\Delta}^{(i)},y))$
(the cost is $N_0Jc_mK$);
then, using these values, for all
$i\in\{1,\ldots,N_0\}$,
$j\in\{1,\ldots,J\}$
and $k\in\{0,1\}^m\setminus\{0\}$,
compute
$\tilde a_{j,k}(X_{\Delta,(j-1)\Delta}^{(i)})$
via~\eqref{eq:coef05a}
(the cost is $N_0J(c_m-1)c_m$).
In this way, the maximal cost order is $JKc_mN_0$.}
$JKc_mN_0$,
which results in the overall cost of order
$JK\max\{KN,c_mN_0\}$.
As for the SRCV algorithm,
we perform $Jc_m$ regressions
with $K$ basis functions in the training phase,
but have in average
$N\PP(\xi_j=y)$ ($\equiv N/c_m$),
$y\in\{-1,1\}^m$,
training paths in each regression,
which again results in the cost of order $JK^2N$,
while in the testing phase we now
have the cost of order $JK(c_m-1)N_0$.
This gives us the overall cost of order
$JK\max\{KN,(c_m-1)N_0\}$,
which is the same order as for the RRCV algorithm.
Finally, regarding the quality of the regressions
in the RRCV and SRCV approaches, it is to expect that
the regressions in the SRCV algorithm,
which are based on the second equality in~\eqref{eq_h},
achieve better approximations
than the regressions in the RRCV algorithm,
provided there are enough training paths
and the basis functions are chosen properly,
because we have
\begin{align}\label{eq:3012a1}
\Var[q_j(X_{\Delta,j\Delta})|X_{\Delta,(j-1)\Delta}=x,\xi_j=y]
=\Var[q_j(\Phi_\Delta(x,y))]=0.
\end{align}
The latter property implies the absence of the statistical  error 
while approximating $h_{j,y}.$
This is well illustrated by the first three plots in Figure~\ref{fig:1}
(the plots are performed for the example
of Section~\ref{subsec:gbm_1}).

\subsection{RCV, RRCV and SRCV algorithms for second order schemes}
\label{sec:23}
Let us define the index set
\begin{align*}
\cI=\left\{(k,l)\in\{1,\ldots,m\}^2:k<l\right\}
\end{align*}
and consider a weak scheme of order~$2$, where
$d$-dimensional approximations
$X_{\Delta,j\Delta}$, $j=0,\ldots,J$, satisfy
$X_{\Delta,0}=x_0$ and
\begin{align}
\label{eq:2002a5}
X_{\Delta,j\Delta}=
\Phi_{\Delta}(X_{\Delta,(j-1)\Delta},\xi_j,V_j),
\quad j=1,\ldots,J,
\end{align}
for some functions
$\Phi_{\Delta}\colon\bbR^{d+m+m(m-1)/2}\to\bbR^d$.
Here,
\begin{itemize}
\item
$\xi_j=(\xi_j^k)_{k=1}^m$, $j=1,\ldots,J$,
are $m$-dimensional random vectors
with i.i.d.\linebreak coordinates satisfying
\begin{align*}
\PP\left(\xi_j^k=\pm\sqrt{3}\right)=\frac16,
\quad
\PP\left(\xi_j^k=0\right)=\frac23,
\end{align*}
\item
$V_j=(V_j^{kl})_{(k,l)\in\cI}$, $j=1,\ldots,J$,
are $m(m-1)/2$-dimensional random vectors
with i.i.d.\ coordinates satisfying
\begin{align*}
\PP\left(V_j^{kl}=\pm1\right)=\frac12,
\end{align*}
\item
the pairs $(\xi_j,V_j)$, $j=1,\ldots,J$, are independent,
\item
for each $j$, the random vectors $\xi_j$ and $V_j$
are independent.
\end{itemize}
An important example of such a scheme
is the
\emph{simplified order~$2$ weak Taylor scheme}
in Section~14.2 of~\cite{KP},
which has the discretisation error
$\EE[f(X_T)]-\EE[f(X_{\Delta,T})]$
of order~$\Delta^2$ under appropriate conditions
on $f$, $\mu$ and~$\sigma$
(also see Theorem~2.1 in~\cite{MilsteinTretyakov:2004}).
Let us introduce the notation
\begin{align*}
\cU=\left\{(o,r)\in\{0,1,2\}^m\times\{0,1\}^\cI:
o_i\ne0\text{ for some }i
\text{ or }
r_{kl}\ne0\text{ for some }k,\;l\right\},
\end{align*}
where $o_i$, $i=1,\ldots,m$
(resp.~$r_{kl}$, $(k,l)\in\cI$),
denote the coordinates of~$o$ (resp.~$r$).
The following result
is an equivalent reformulation
of Theorem~3.5 in~\cite{belomestny2016variance}.

\begin{theorem}
\label{th:weak_md03}
The following representation holds
\begin{align}
\label{eq:2002a1}
f(X_{\Delta,T})=\EE f(X_{\Delta,T})+
\sum_{j=1}^J
\sum_{(o,r)\in\cU}
a_{j,o,r}(X_{\Delta,(j-1)\Delta})
\prod_{i=1}^m H_{o_i}(\xi_j^i)
\prod_{(k,l)\in\cI}(V_j^{kl})^{r_{kl}},
\end{align}
where $H_0(x)\doteq1$,
$H_1(x)\doteq x$, $H_2(x)\doteq\frac{x^2-1}{\sqrt{2}}$,
and the coefficients
$a_{j,o,r}\colon\bbR^d\to\bbR$
are given by the formula
\begin{align}
\label{eq:2002a2}
a_{j,o,r}(x)
=\EE\left[\left.
f(X_{\Delta,T})
\prod_{i=1}^m H_{o_i}(\xi_j^i)
\prod_{(k,l)\in\cI}(V_j^{kl})^{r_{kl}}
\right| X_{\Delta,(j-1)\Delta}=x
\right]
\end{align}
for all $j\in\{1,\ldots,J\}$ and $(o,r)\in\cU$.
\end{theorem}

Thus, with
\begin{align}\label{eq:2512b1}
M^{(2)}_{\Delta,T}\doteq
\sum_{j=1}^J
\sum_{(o,r)\in\cU}
a_{j,o,r}(X_{\Delta,(j-1)\Delta})
\prod_{i=1}^m H_{o_i}(\xi_j^i)
\prod_{(k,l)\in\cI}(V_j^{kl})^{r_{kl}},
\end{align}
we have $\EE\left[M^{(2)}_{\Delta,T}\right]=0$
and $\Var\left[f(X_{\Delta,T}) - M^{(2)}_{\Delta,T}\right]=0$
in the case of second order schemes.
The RCV approach for second order schemes
relies on Theorem~\ref{th:weak_md03}
in the same way as the one for first order schemes
relies on Theorem~\ref{th:weak_md01}.

We now introduce the functions $q_j(x)$,
$j\in\{1,\ldots,J\}$, by formula~\eqref{tower:q}
also in the case of second order schemes and,
for all $y\in\{-\sqrt3,0,\sqrt3\}^m$, set
\begin{align}\label{eq:3112a1}
p_m(y)\doteq
\frac{4^{\sum_{i=1}^m I(y_i=0)}}{6^m 2^{\frac{m(m-1)}2}}.
\end{align}
Notice that $p_m(y)=\PP(\xi_j=y,V_j=z)$
for all $z\in\{-1,1\}^\cI$.
The next result is Proposition~3.7
of~\cite{belomestny2016variance}.

\begin{proposition}
\label{prop:2812a1}
We have $q_J\equiv f$ and, for each $j\in\{2,\ldots,J\}$,
\begin{align}
\label{eq:2812a2}
q_{j-1}(x)=&\EE\bigl[q_j(X_{\Delta, j\Delta})|X_{\Delta,(j-1)\Delta}=x\bigr]=\!\!\!\!
\sum_{y\in\{-\sqrt3,0,\sqrt3\}^m}\;
\sum_{z\in\{-1,1\}^\cI}
p_m(y)\,q_j(\Phi_\Delta(x,y,z)).
\end{align}
Moreover, for all $j\in\{1,\ldots,J\}$ and $(o,r)\in\cU$,
the functions $a_{j,o,r}(x)$ of~\eqref{eq:2002a2}
can be expressed in terms of the functions
$q_j(x)$ as
\begin{align}\label{eq:2512b2}
a_{j,o,r}(x)=\!\!\!\!
\sum_{y\in\{-\sqrt3,0,\sqrt3\}^m}\;
\sum_{z\in\{-1,1\}^\cI}
\left[
p_m(y)
\prod_{i=1}^m H_{o_i}(y_i)
\prod_{(k,l)\in\cI} z_{kl}^{r_{kl}}
\right] q_j(\Phi_\Delta(x,y,z)),
\end{align}
where $o_i$ and $y_i$, $i=1,\ldots,m$,
denote the coordinates of $o$ and~$y$,
while $r_{kl}$ and $z_{kl}$, $(k,l)\in\cI$,
are the coordinates of $r$ and~$z$.
\end{proposition}
Similar to~\eqref{eq_h},
we define functions $h_{j,y,z}$,
for all $j\in\{1,\ldots,J\}$, $y\in\{-\sqrt3,0,\sqrt3\}^m$
and $z\in\{-1,1\}^\cI$, by the formula
\begin{align}\label{eq_h2}
h_{j,y,z}(x)\doteq q_j(\Phi_\Delta(x,y,z))
=\EE[q_j(X_{\Delta,j\Delta})|X_{\Delta,(j-1)\Delta}=x,\xi_j=y,V_j=z].
\end{align}
The RRCV and SRCV algorithms for second order schemes
now rely on Proposition~\ref{prop:2812a1} and on~\eqref{eq_h2}
in the same way as the ones for first order schemes
rely on Proposition~\ref{prop:2202a1} and on~\eqref{eq_h}.
The whole discussion in the end of Section~\ref{sec:22},
and, in particular, the formula $JK\max\{KN,(c_m-1)N_0\}$
for the overall cost order of the SRCV algorithm,
apply also in the case of second order schemes,
where we only need to change the value of~$c_m$:
here $c_m\doteq3^m 2^{m(m-1)/2}$.

\section{Complexity analysis}
\label{sec:3}
In this section we extend
the complexity analysis presented
in~\cite{belomestny2016variance}
to the case of the stratified regression algorithm.
Below we only sketch the main results
for the second order schemes.
We make the following assumptions.
\begin{itemize}
\item[(A1)]
All functions  \(h_{j,y,z}(x)\) of~\eqref{eq_h2}
are uniformly bounded, i.e.\ there is a constant \(A>0\)
such that
$\sup_{x\in\bbR^d} |h_{j,y,z}(x)|\le A<\infty$.

\item[(A2)]
The functions \(h_{j,y,z}\left(x\right)\) can be
well approximated by the functions from
$\Psi_{K}\doteq\text{span}\left(\left\{\psi_{1},\ldots,\psi_{K}\right\}\right)$,
in the sense that there are constants
$\kappa>0$ and $C_\kappa>0$ such that
\begin{align*}
\inf_{g\in\Psi_{K}}\int_{\mathbb{R}^d}\left(h_{j,y,z}\left(x
\right)-g\left(x\right)\right)^2\,\PP_{\Delta,j-1}(dx)\leq  \frac{C_\kappa}{K^{\kappa}},
\end{align*}
where $\PP_{\Delta,j-1}$
denotes the distribution of~$X_{\Delta,(j-1)\Delta}$.
\end{itemize}

\begin{remark}\label{rem:1301a1}
A sufficient condition for~(A1) is boundedness of~$f$.
As for~(A2),
this is a natural condition to be satisfied
for good choices of $\Psi_K$.
For instance, under appropriate assumptions,
in the case of
\emph{piecewise polynomial regression}
as described in~\cite{belomestny2016variance},
(A2)~is satisfied with
$\kappa=\frac{2\nu(p+1)}{2d(p+1)+d\nu}$,
where the parameters $p$ and $\nu$
are explained in~\cite{belomestny2016variance}.
\end{remark}

In Lemma~\ref{th:2104a1}
below we present an $L^2$-upper bound
for the estimation error on step~2
of the training phase of the SRCV algorithm
(see page~\pageref{step2}).
To this end, we need to describe more precisely,
how exactly the regression-based approximations
$\tilde h_{j,y,z}$ are constructed:
\begin{itemize}
\item[(A3)]
Let functions $\hat h_{j,y,z}(x)$
be obtained by linear regression
(based on the second equality in~\eqref{eq_h2})
onto the set of basis functions
$\left\{\psi_{1},\ldots,\psi_{K}\right\}$,
while the approximations
$\tilde h_{j,y,z}(x)$ on step~2
of the training phase of the SRCV algorithm
be the truncated estimates, which are defined as follows:
\begin{align*}
\tilde h_{j,y,z}(x)\doteq
T_A\hat h_{j,y,z}(x)\doteq
\begin{cases}
\hat h_{j,y,z}(x)&\text{if }|\hat h_{j,y,z}(x)|\le A,\\
A\sgn\hat h_{j,y,z}(x)&\text{otherwise}
\end{cases}
\end{align*}
($A$ is the constant from~(A1)).
\end{itemize}

\begin{lemma}\label{th:2104a1}
Under~(A1)--(A3), we have
\begin{align}\label{eq:2104a2}
\E\|\tilde h_{j,y,z}-h_{j,y,z}\|^2_{L^2(\PP_{\Delta,j-1})}
\le\tilde{c}\, A^2(\log N+1)\frac{K}{Np_m(y)}
+\frac{8\,C_\kappa}{K^\kappa},
\end{align}
where $\tilde{c}$ is a universal constant and $p_m(y)$
is given in~\eqref{eq:3112a1}.
\end{lemma}

It is necessary to explain once in detail
how to understand the left-hand side of~\eqref{eq:2104a2}.
The functions $\hat h_{j,y,z}(x)$ (see~(A3))
are linear combinations
$\alpha_1\psi_1(x)+\ldots+\alpha_K\psi_K(x)$
of the basis functions,
where the coefficients $\alpha_i$
are random in that they depend
on the simulated training paths.
That is, we have, in fact,
$\hat h_{j,y,z}(x)=\hat h_{j,y,z}(x,D_N^{tr})$
and, consequently,
$\tilde h_{j,y,z}(x)=\tilde h_{j,y,z}(x,D_N^{tr})$
(cf.~\eqref{eq:2712a2}).
Thus, the expectation in the left-hand side of~\eqref{eq:2104a2}
means averaging over the randomness in~$D_N^{tr}$.

The next step is to provide
an upper bound for the regression-based
estimates of the coefficients~$a_{j,o,r}$,
which are constructed on step~3
of the training phase of the SRCV algorithm.

\begin{lemma}\label{th:2105a1}
Under~(A1)--(A3), we have
\begin{align}\label{eq:2105a2}
\E\|\tilde a_{j,o,r}-a_{j,o,r}\|^2_{L^2(\PP_{\Delta,j-1})}\le c_m\tilde{c}\, A^2(\log N+1)\frac{K}{N}
+\frac{8\,C_\kappa}{K^\kappa}C_{m,o},
\end{align}
where $C_{m,o}\doteq\sum_{y\in\{-\sqrt3,0,\sqrt3\}^m}\;
c_m2^\frac{m(m-1)}{2}\left[
p_m(y)\prod_{i=1}^m H_{o_i}(y_i)
\right]^2$.
\end{lemma}

Let $(X_{\Delta,j\Delta})_{j=0,\ldots,J}$
be a testing path, which is independent of the
training paths $D_N^{tr}$.
We now define
\begin{align}\label{eq:0101a1}
\tilde M^{(2)}_{\Delta,T}\doteq
\sum_{j=1}^J
\sum_{(o,r)\in\cU}
\tilde a_{j,o,r}(X_{\Delta,(j-1)\Delta},D_N^{tr})
\prod_{i=1}^m H_{o_i}(\xi_j^i)
\prod_{(k,l)\in\cI}(V_j^{kl})^{r_{kl}}
\end{align}
(cf.~\eqref{eq:2512b1})
and bound the variance
$\Var[f(X_{\Delta,T})-\tilde M_{\Delta,T}^{(2)}]$
from above.\footnote{\label{ft:1301a1}Notice that the variance
of the SRCV estimate
$\frac{1}{N_0}\sum_{i=1}^{N_0}
\left[f\Bigl(X_{\Delta,T}^{(i)}\Bigr)-\tilde M_{\Delta,T}^{(2),(i)}\right]$
with $N_0$ testing paths
is $\frac1{N_0}\Var[f(X_{\Delta,T})-\tilde M_{\Delta,T}^{(2)}]$.}
With the help of Lemmas~\ref{th:2104a1} and~\ref{th:2105a1}
we now derive the main result of this section:

\begin{theorem}
\label{th:2106a1}
Under~(A1)--(A3), it holds
\begin{align*}
\Var[f(X_{\Delta,T})-
\tilde{M}^{(2)}_{\Delta,T}]
\le J\left(\left(c_m-1\right)c_m\tilde{c}\, A^2(\log N+1)\frac{K}{N}
+\frac{8\,C_\kappa}{K^\kappa}\tilde c_m\right),
\end{align*}
where $\tilde c_m=c_m-\left(\frac32\right)^m$.
\end{theorem}

The preceding theorem allows us
to perform complexity analysis for the SRCV approach,
which means that we want to find the minimal
order of the overall computational cost
necessary to implement the algorithm
under the constraint that the
mean squared error is of order~$\varepsilon^2$.
The overall cost is of order $JK\max\left\{NK,(c_m-1)N_0\right\}$.
We have the constraint
\begin{equation}\label{eq:19022017a1}
\EE\left[\left(V_{N_0}^{SRCV}-\EE f(X_T)\right)^2\right]\lesssim\varepsilon^2,
\end{equation}
where
$$
V_{N_0}^{SRCV}:=\frac{1}{N_0}\sum_{i=1}^{N_0}
\left[f\Bigl(X_{\Delta,T}^{(i)}\Bigr)-\tilde M_{\Delta,T}^{(2),(i)}\right].
$$
Since
\begin{equation}\label{eq:19022017a2}
\EE\left[\left(V_{N_0}^{SRCV}-\EE f(X_T)\right)^2\right]
=\left(\EE f(X_{\Delta,T})-\EE f(X_T)\right)^2
+\frac1{N_0}\Var\left[f(X_{\Delta,T})-\tilde M_{\Delta,T}^{(2)}\right],
\end{equation}
constraint~\eqref{eq:19022017a1} reduces to
$$
\max\left\{\frac{1}{J^4},
\frac{JK\log(N)c_m(c_m-1)}{NN_0},
\frac{JC_\kappa\tilde c_m}{K^\kappa N_0}\right\}\lesssim\varepsilon^2,
$$
where the first term comes from the squared bias of the estimator
(see the first term in the right-hand side of~\eqref{eq:19022017a2})
and the remaining two ones come from the variance of the estimator
(see the second term in the right-hand side of~\eqref{eq:19022017a2} and apply Theorem~\ref{th:2106a1}).
It is natural to expect that the optimal solution is given by all constraints being active  as well as $NK\asymp (c_m-1) N_0$,
that is,
both terms in the overall cost are of the same order.
Provided that\footnote{Performing
the full complexity analysis via Lagrange multipliers one can see that these parameter values are
\emph{not} optimal if $\kappa\le 1$
(a Lagrange multiplier corresponding to
a ``$\le0$'' constraint is negative).
Recall that in the case of piecewise polynomial regression
(see~\cite{belomestny2016variance}
and recall Remark~\ref{rem:1301a1})
we have $\kappa=\frac{2\nu(p+1)}{2d(p+1)+d\nu}$.
Let us note that
in~\cite{belomestny2016variance} it is required
to choose the parameters $p$ and $\nu$ according to
$p>\frac{d-2}2$ and $\nu>\frac{2d(p+1)}{2(p+1)-d}$,
which implies that $\kappa>1$,
for $\kappa$ expressed via $p$ and $\nu$ by the above formula.}
$\kappa>1$, we obtain the following parameter values:
\begin{gather*}
J\asymp\varepsilon^{-\frac{1}{2}},\quad K\asymp \left[\frac{\tilde c_m^2C_\kappa^2}{c_m\varepsilon^{\frac{5}{2}}}\right]^\frac{1}{2\kappa+2},\quad N\asymp \frac{(c_m-1)\sqrt{c_m}}{\varepsilon^\frac{5}{4}}\sqrt{\log\left(\varepsilon^{-\frac{5}{4}}\right)},
\\ 
N_0\asymp\frac{NK}{c_m-1}\asymp \left[\frac{c_m^\kappa \tilde c_m^2C_\kappa^2}{\varepsilon^{\frac{5\kappa+10}{2}}}\right]^\frac{1}{2\kappa+2}\sqrt{\log\left(\varepsilon^{-\frac{5}{4}}\right)}.
\end{gather*}
Thus, we have for the complexity
\begin{align}
\label{eq:complt}
\mathcal{C}&\asymp JNK^2 \asymp JN_0K(c_m-1)\asymp \left[\frac{(c_m-1)^{2\kappa+2}c_m^{\kappa-1} \tilde c_m^4C_\kappa^4}{\varepsilon^{\frac{7\kappa+17}{2}}}\right]^\frac{1}{2\kappa+2}\sqrt{\log\left(\varepsilon^{-\frac{5}{4}}\right)}. 
\end{align}
Note that the $\log$-term in the solution of $N$ and $N_0$ has been added afterwards to satisfy all constraints. Complexity estimate~\eqref{eq:complt}
shows that one can go beyond the complexity order
$\varepsilon^{-2}$, provided that $\kappa>9$,
and that we can achieve the complexity order
$\varepsilon^{-1.75-\delta}$,
for arbitrarily small $\delta>0$,
provided $\kappa$ is large enough.

\section{Numerical results}
\label{sec:4}
In this section, we present several numerical examples showing the efficiency of the SRCV approach. It turns that even the weak Euler scheme~\eqref{eq:PhiK=1}
already shows the advantage of the new  methodology over the standard Monte Carlo (SMC) as well as over the original RCV and RRCV approaches in terms of variance reduction effect. 
Regarding the choice of basis functions,
we use for the RCV, RRCV and SRCV approaches
polynomials of degree $\le p$, that is,
$\psi_l(x)=\prod_{i=1}^dx_i^{l_i}$,
where $l=(l_1,\ldots l_d)\in\left\{0,1,\ldots,p\right\}^d$
and $\sum_{l=1}^dl_i\leq p$.
In addition to the polynomials,
we consider the function $f$ as a basis function.
We choose $J=100$, $N=10^5$, $N_0=10^7$, $p=1$ in all examples. Hence, we have overall $K=\binom{p+d}{d}+1=d+2$ basis functions in each regression. 
Then we compute the estimated variances for the
SMC, RCV, RRCV and SRCV approaches.
More precisely, when speaking about
``variance'' below
(e.g.\ in Tables \ref{tab_1d_linz},
\ref{tab_10d_gbm} and~\ref{tab_9d_heston})
we mean sample variance of one summand
$f(X_{\Delta,T}^{(i)})-\tilde M_{\Delta,T}^{(1),(i)}$
(see~\eqref{eq:2812b0})
in the case of RCV, RRCV and SRCV,
while, in the case of SMC,
the sample variance of
$f(X_{\Delta,T}^{(i)})$ is meant.
Thus, we analyse the variance reduction effect only, since the bias is the same for all these methods.
To measure the numerical performance
of a variance reduction method,
we look at the ratio of variance vs.\ computational time,
i.e., for the SRCV, we look at
\begin{align*}
\theta_{\text{SRCV}}\doteq\frac{\textrm{Var}_{\text{SRCV}}}{\textrm{Var}_{\text{SMC}}}\cdot\frac{\textrm{Time}_{\text{SRCV}}}{\textrm{Time}_{\text{SMC}}},
\end{align*}
where $\textrm{Var}_{\text{SRCV}}$ and $\textrm{Time}_{\text{SRCV}}$ denote the variance and
the overall computational time of the SRCV approach
($\textrm{Var}_{\text{SMC}}$ and
$\textrm{Time}_{\text{SMC}}$ have the similar meaning).
The smaller $\theta_{\text{SRCV}}$ is, the more profitable is the SRCV algorithm compared to the SMC one.
We similarly define $\theta_{\text{RCV}}$
and $\theta_{\text{RRCV}}$
(each of the regression-based algorithms
is compared with the SMC approach).

\subsection{Geometric Brownian motion (GBM) with high volatility}
\label{subsec:gbm_1}
\begin{figure}[htbp!]
\centering
\includegraphics[width=0.49\textwidth]{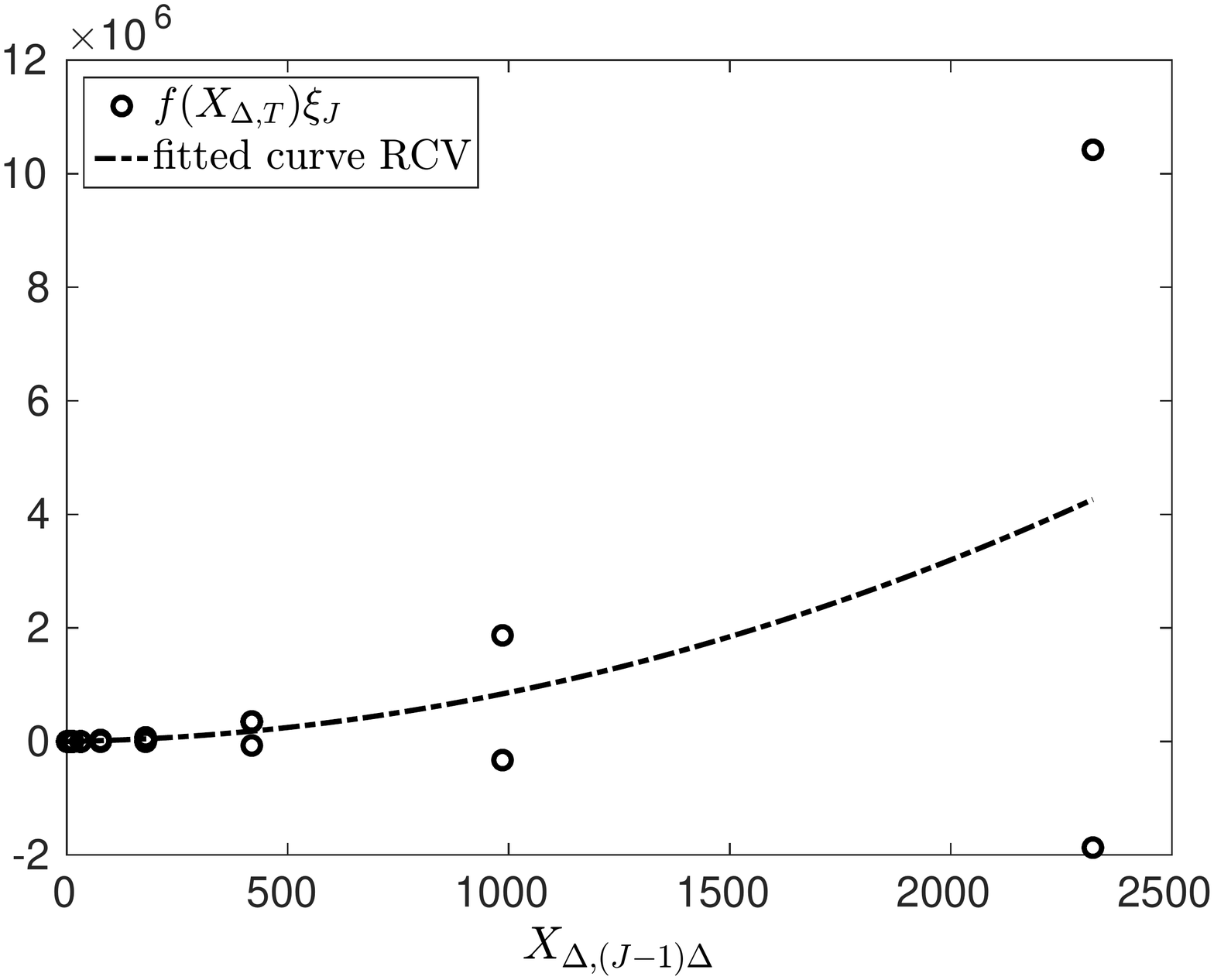}
\includegraphics[width=0.49\textwidth]{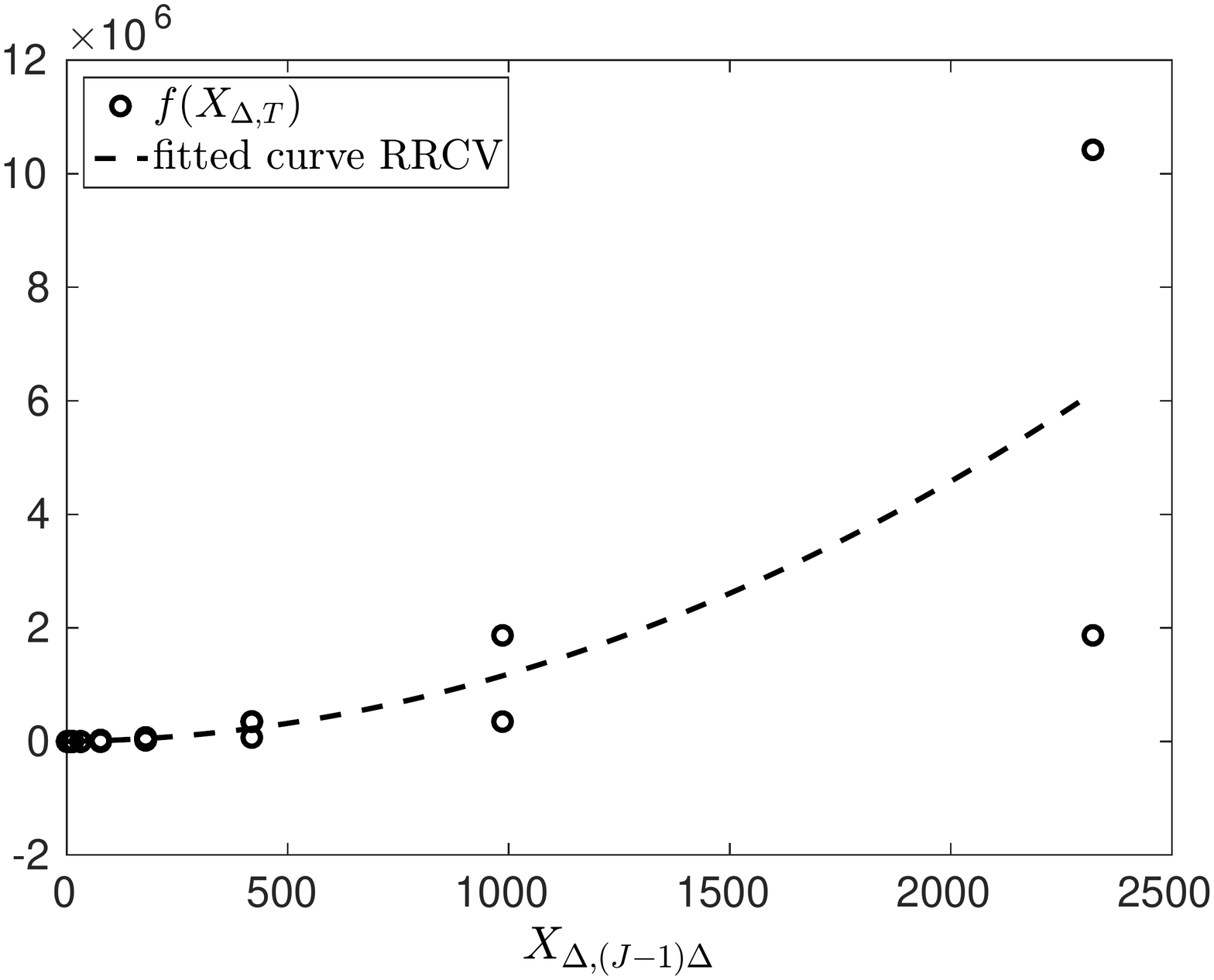}
\includegraphics[width=0.49\textwidth]{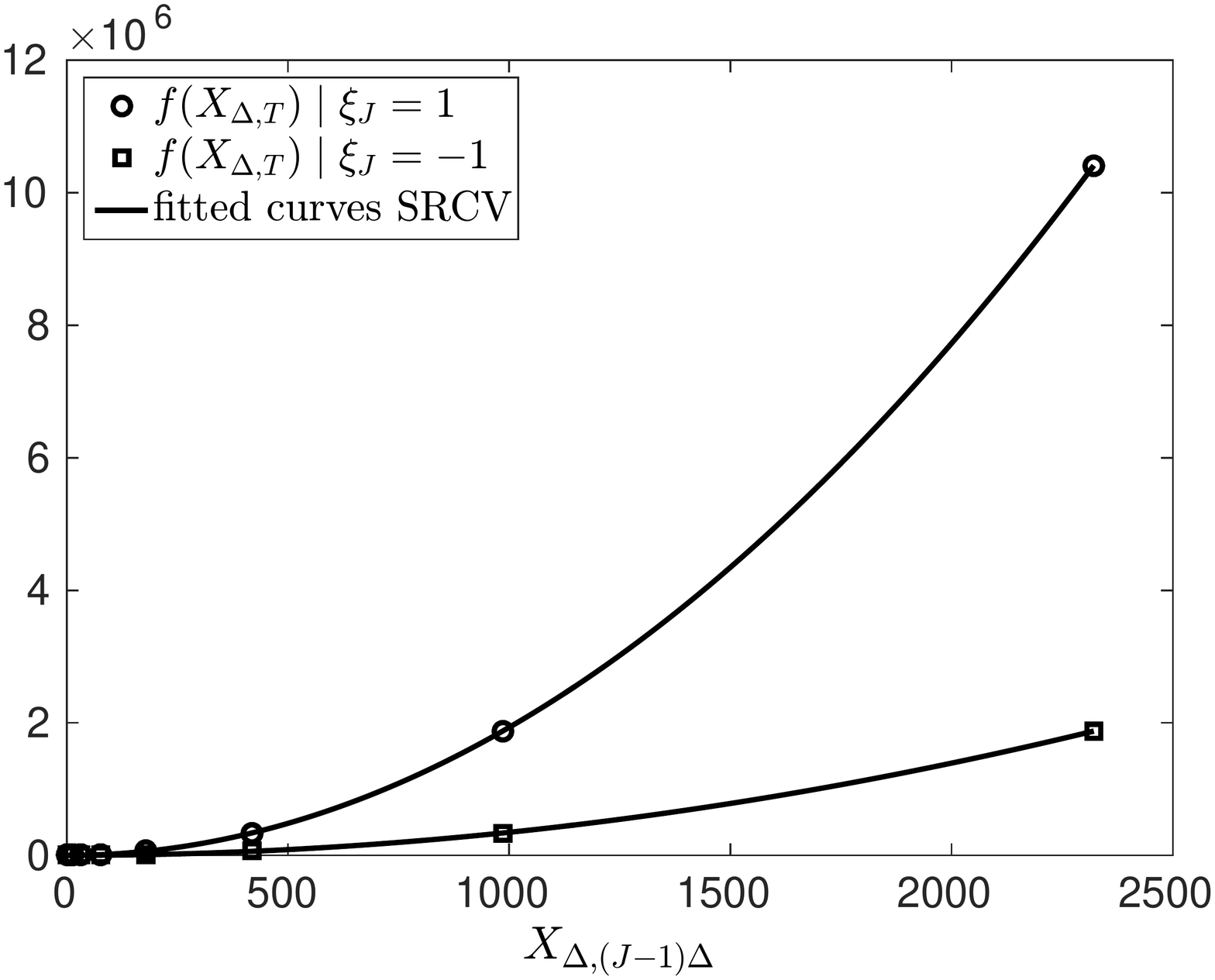}
\includegraphics[width=0.49\textwidth]{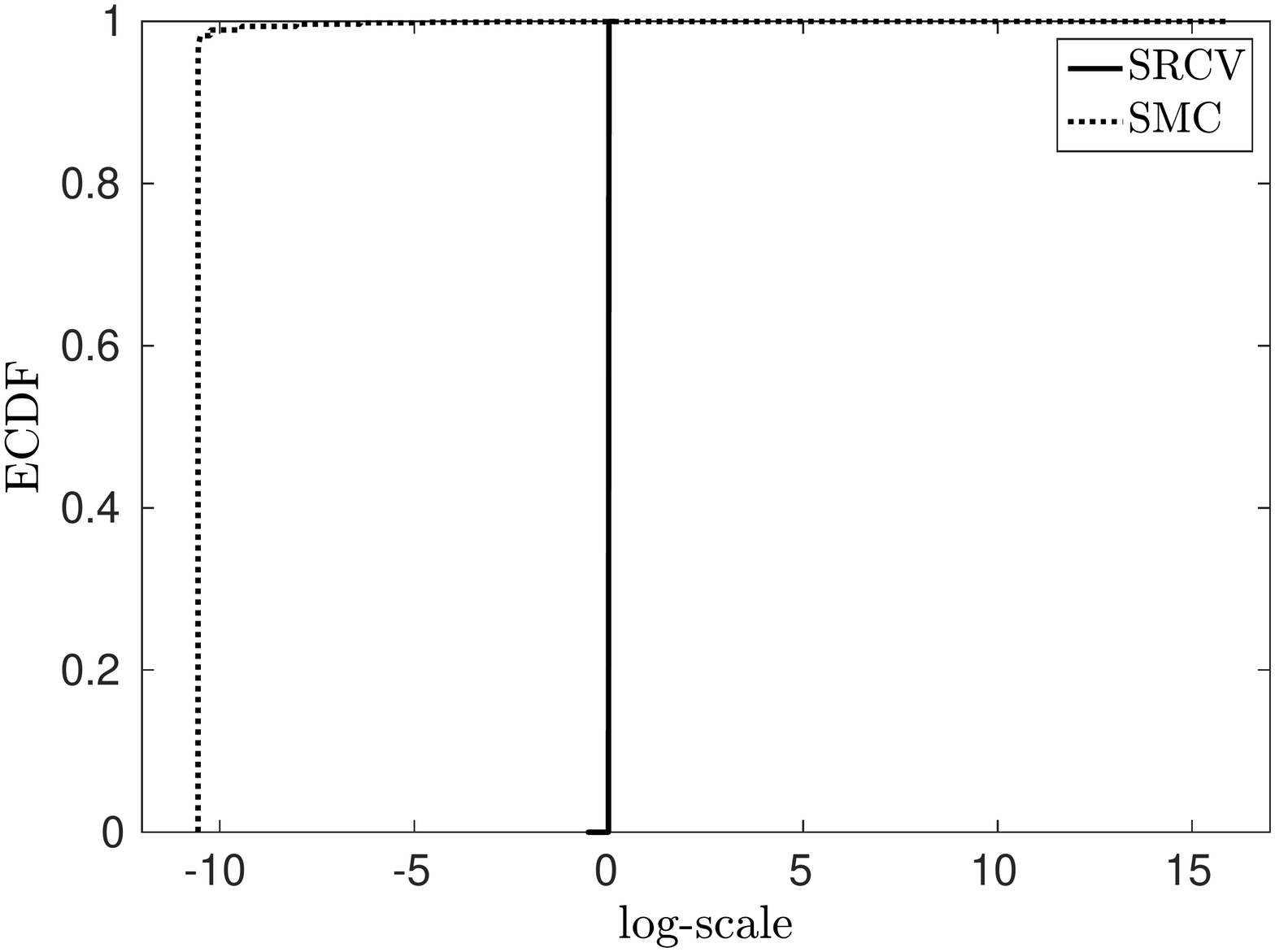}
\includegraphics[width=0.49\textwidth]{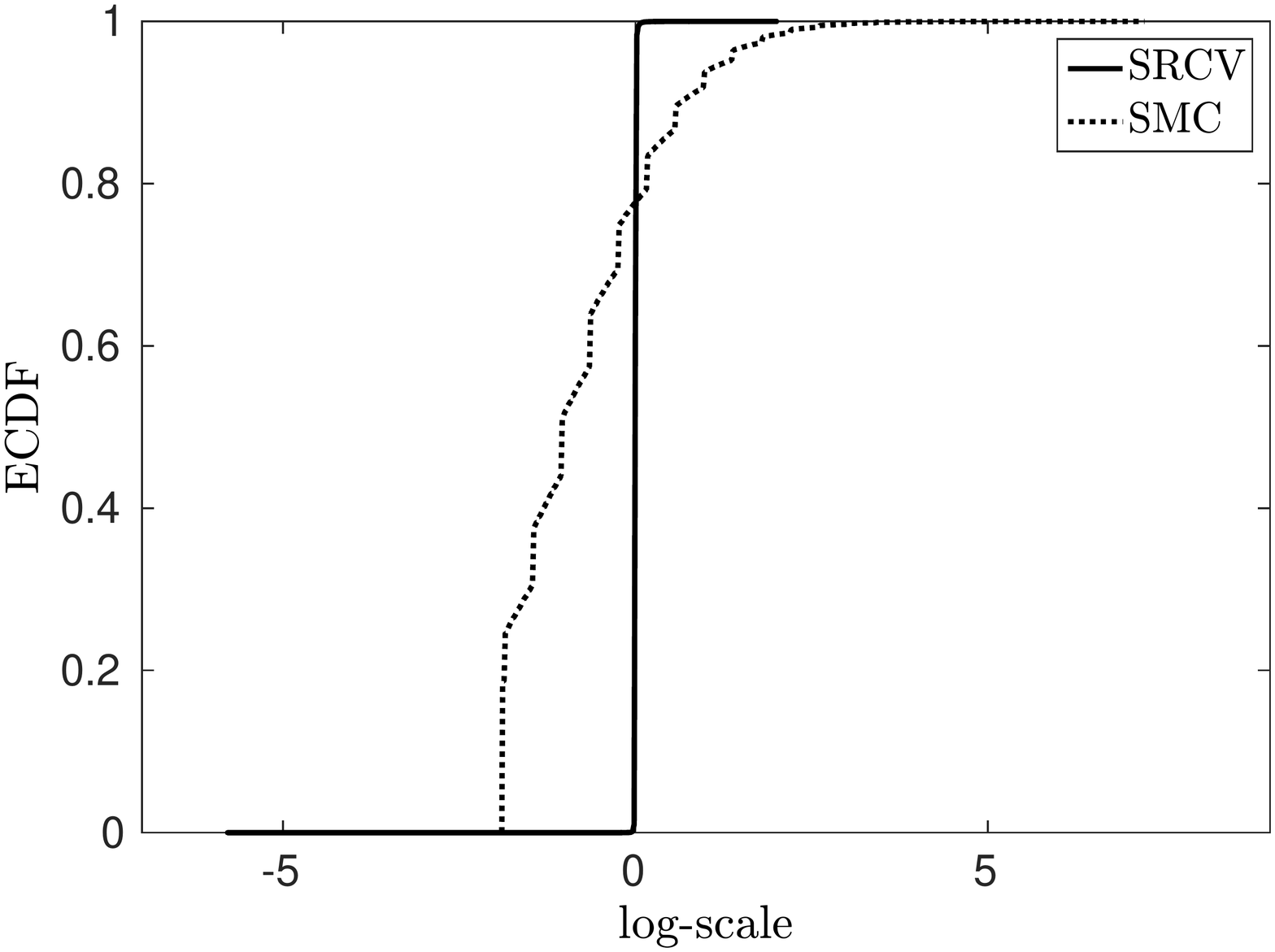}
\includegraphics[width=0.49\textwidth]{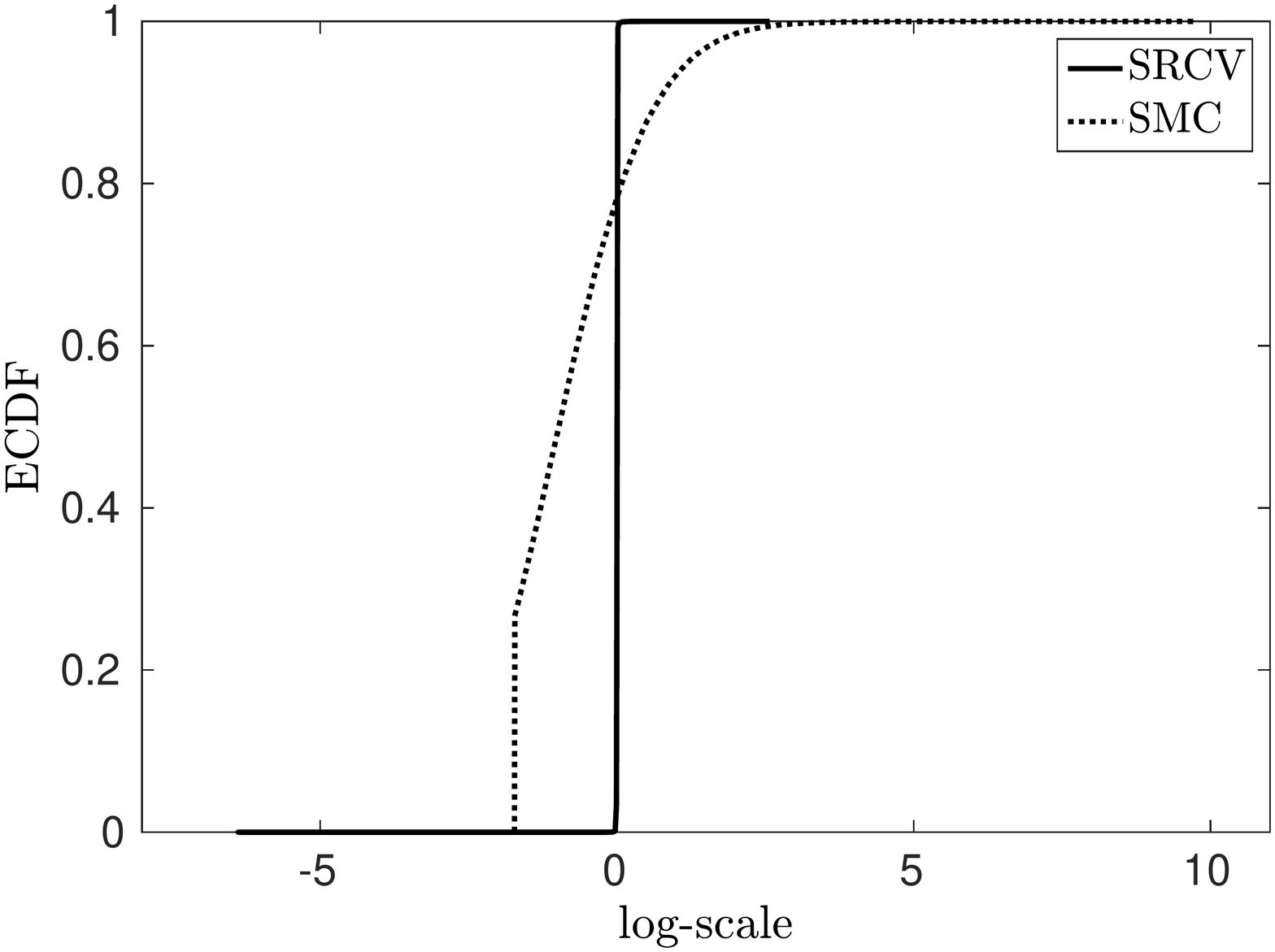}
\captionsetup{singlelinecheck=off,margin=1cm}
\caption[]{
\begin{enumerate}[leftmargin=*]
\item top left: first regression task for RCV (Section \ref{subsec:gbm_1}),
\item top right: first regression task for RRCV (Section \ref{subsec:gbm_1}),
\item center left: first regression task for SRCV (Section \ref{subsec:gbm_1}),
\item center right: ECDF of the log-scaled sample for SRCV and SMC (Section \ref{subsec:gbm_1}),
\item bottom left: ECDF of the log-scaled sample for SRCV and SMC (Section \ref{subsec:gbm_10}),
\item bottom right: ECDF of the log-scaled sample for SRCV and SMC (Section \ref{subsec:heston_9}).
\end{enumerate}
}
\label{fig:1}
\end{figure}

Here $d=m=1$ ($K=3$).
We consider the following SDE
\begin{align}\label{eq:sde1}
dX_t=&rX_tdt+
\sigma X_tdW_t,\quad X_0=1,
\end{align}
for $t\in\left[0,1\right]$, where $r=-1$ and $\sigma=4$. 
Furthermore, we consider the functional
$f(x)=x^2$.
In the following, we plot the empirical cumulative distribution function (ECDF)
of the ``log-scaled sample'', which is
\begin{align*}
\log(1+f_i-f_{\min})-\log(1+\bar{f}-f_{\min})
\end{align*}
for the SMC, and
\begin{align*}
\log(1+u_i-u_{\min})-\log(1+\bar{u}-u_{\min})
\end{align*}
for the RCV, and RRCV and SRCV, where 
\begin{align*}
f_i&\doteq f(X_{\Delta,T}^{(i)}),\quad u_i\doteq f_i-\tilde{M}_{\Delta,T}^{(1),(i)},\quad i\in\left\{1,\ldots,N_0\right\},\\
f_{\min}&\doteq \min_{i=1,\ldots,N_0}f_i,\quad u_{\min}\doteq \min_{i=1,\ldots,N_0}u_i,\quad \bar{f}\doteq \frac1{N_0}\sum_{i=1}^{N_0}f_i, \quad
\bar{u}\doteq \frac1{N_0}\sum_{i=1}^{N_0}u_i.
\end{align*}
The results for such a log-scaled sample
are illustrated in Table~\ref{tab_1d_linz}.
As can be also seen from the fourth plot in Figure~\ref{fig:1}
(ECDFs of the SRCV and SMC),
the variance reduction works absolutely fine for SRCV.
Most of the sample values produced by SMC are much smaller than the corresponding mean value,
whereas the deviation w.r.t.\ the mean $\bar{u}$ is very small for the SRCV approach. The main problem of the SMC approach in this case is that almost all paths tend to zero so that the small number of outliers is not sufficient to reach the (large)
expectation $\EE[f(X_{\Delta,T})]$, i.e.\ $N_0$ has to be increased a lot to approach the expectation. In contrast, for the SRCV approach all paths (paths close to zero as well as outliers) are ``shifted'' close to the expectation and thus we obtain a very small variance.
We only plot the ECDFs of the SRCV and SMC in Figure~\ref{fig:1}, since the ECDFs of the RCV and RRCV look visually very similar to that for SRCV.
The difference is, however, revealed
in the ``Min'' and ``Max'' columns
of the Table~\ref{tab_1d_linz}.
That is, the RCV and RRCV algorithms produce several outliers
which result in that the RCV and RRCV do not give us any variance reduction effect!
One reason for this significant difference between the algorithms
is given in the first three plots in Figure~\ref{fig:1}, where we illustrate the regression results for the RCV, RRCV and SRCV algorithms at the last time point,
which means the first regression task.
Here, we have accurate estimates only for the SRCV
(cf.~the discussion around~\eqref{eq:3012a1}).
\begin{table}[htb!]
\small
\centering
\begin{tabular}{|l|l|l|l|l|l|}
\hline
Approach & \multicolumn{1}{l|}{Min} & \multicolumn{1}{l|}{Max} & \multicolumn{1}{l|}{Variance} & \multicolumn{1}{l|}{Time (sec)} & \multicolumn{1}{l|}{$\theta$} \\ \hline
SRCV & -0.5 & 0.2 & $6.3\cdot 10^{-8}$ & 30.5 & $1.32\cdot 10^{-23}$ \\ \hline
RRCV & -25.4 & 1.7 & $2.7\cdot 10^{16}$ & 65.3 & 12.38 \\ \hline
RCV & -27.8 & 0.1 & $1.4\cdot 10^{17}$ & 30.0 & 28.57 \\ \hline
SMC & -10.6 & 15.9 & $9.6\cdot 10^{15}$ & 15.1 & 1\\ \hline
\end{tabular}
\caption{Results of the algorithms for a quadratic function $f$ under a GBM model.}
\label{tab_1d_linz}
\end{table}

\subsection{High-dimensional geometric Brownian motion}
\label{subsec:gbm_10}
We consider the following SDE for $d=m=10$ ($K=12$):
\begin{align*}
dX_t^i=rX_t^idt+\sigma^iX_t^iA^idW_t,\quad t\in\left[0,1\right],\quad i=1,\ldots,10,
\end{align*}
where $X_0^i=1$, $\sigma^i=2\,\forall i,$ $r=0.05$ and $A^i\doteq\begin{pmatrix}
A^{i,1} \cdots A^{i,10}\end{pmatrix}$, $AA^T=\left(\rho_{ik}\right)_{i,k=1,\ldots,10}$ with $\rho_{ik}=\rho_{ki}\in\left[-1,1\right]$ and $\rho_{ik}=1$ for $i=k$ (that is,
$A^i W$, $i=1,\ldots,10$, are correlated Brownian motions).
For $i<k$ we choose 
\begin{align*}
\rho_{ik}=\left\{\begin{array}{lllrll}
0.9 & \text{if }i=1,\,k=2, && -0.95 & \text{if }i=3,\,k=4,\\
0.5 & \text{if }i=5,\,k=6, && -0.9 & \text{if }i=7,\,k=8,\\
0.8 & \text{if }i=9,\,k=10, && 0 & \text{otherwise}.\end{array}\right.
\end{align*}
In this example, we illustrate the performances
of the algorithms by means of the functional
$f\left(x\right)=\max\left\{\max_{i\in\left\{1,\ldots,10\right\}}x^i-1,0\right\}$.
For saving a lot of computing time, we use the ``simplified control variate'' 
\begin{align*}
\tilde{\tilde{M}}^{(1)}_{\Delta,T}\doteq
\sum_{j=1}^J
\sum_{r=1}^m
\tilde a_{j,r}(X_{\Delta,(j-1)\Delta},D_N^{tr})\xi_j^r
\end{align*}
rather than $\tilde M_{\Delta,T}^{(1)}$ for RCV and SRCV,
where $\tilde a_{j,r}$ is a shorthand notation
for $\tilde a_{j,k(r)}$ with $k(r)=(k_1,\ldots,k_m)$,
$k_r=1$, $k_i=0$ for $i\ne r$.
This simplification already takes
much of the variance reduction power into account,
while significantly reduces the number of summands
needed to construct the control variate
($m=10$ vs.\ $c_m-1=2^m-1=1023$ summands
in the second sum above).
For the SRCV algorithm, this results in the cost order
$N_0JmK$ instead of $N_0J(c_m-1)K$
in the testing phase
($10^{11}$ vs.\ $10^{13}$ in this example).
Such a reduction in computational time
due to using $\tilde{\tilde{M}}^{(1)}_{\Delta,T}$
applies also to the RCV algorithm,
but does not apply to the RRCV algorithm.
Namely, with $\tilde{\tilde{M}}^{(1)}_{\Delta,T}$
the testing phase of the RRCV algorithm
would now cost $N_0Jc_mK+N_0Jmc_m$
(in the second summand we now have the factor $m$
instead of $c_m-1$,
cf.\ footnote~\ref{ft:3013a1} on page~\pageref{ft:3013a1}),
which is still of order $10^{13}$ in the present example.
Therefore, we do not consider the RRCV
approach in this example.
The results for the log-scaled sample are illustrated in Table~\ref{tab_10d_gbm}. Again, the SRCV approach achieves a much smaller variance compared to the SMC and RCV
(see the fifth plot in Figure~\ref{fig:1}).
\begin{table}[htb!]
\small
\centering
\begin{tabular}{|l|l|l|l|l|l|}
\hline
Approach & \multicolumn{1}{l|}{Min} & \multicolumn{1}{l|}{Max} & \multicolumn{1}{l|}{Variance} & \multicolumn{1}{l|}{Time (sec)} & \multicolumn{1}{l|}{$\theta$}  \\ \hline
SRCV & -5.8 & 2.0 & 14.6 & 573.9 & 0.13 \\ \hline
RCV & -10.4 & 0.7 & 11271.0 & 288.2 & 51.50 \\ \hline
SMC & -1.9 & 7.2 & 448.9 & 140.5 & 1 \\ \hline
\end{tabular}
\caption{Results of the algorithms for a Call-on-max-option under a high-dimensional GBM.}
\label{tab_10d_gbm}
\end{table}

\subsection{High-dimensional Heston model}
\label{subsec:heston_9}
We consider the following SDE for $d=m=9$ ($K=11$):
\begin{align*}
dX_t^i&=rX_t^idt+\sigma^iX_t^i\sqrt{X_t^9}A^idW_t,
\quad i=1,\ldots,8,\\
dX_t^9&=\lambda\left(\bar{v}-X_t^9\right)dt+\eta\sqrt{X_t^9}A^9dW_t,
\end{align*}
where $t\in\left[0,1\right]$, $X_0^i=1$, $\sigma^i=1$ for $i=1,\ldots,8$ as well as $X_0^9=4$, $r=0.05$, $\lambda=0.1$, $\bar{v}=4$, $\eta=1$ and $A^i\doteq\begin{pmatrix}
A^{i,1} \cdots A^{i,9}\end{pmatrix}$, $AA^T=\left(\rho_{ik}\right)_{i,k=1,\ldots,9}$.
Here, for $i<k$ we choose
\begin{align*}
\rho_{ik}=\left\{\begin{array}{lllrll}
0.9 & \text{if }i=1,\,k=2, && -0.95 & \text{if }i=3,\,k=4,\\
0.5 & \text{if }i=5,\,k=6, && -0.9 & \text{if }i=7,\,k=8,\\
-0.2 & \text{if }i\in\left\{1,2,3,5,6,7\right\},\,k=9, &&
0.2 & \text{if }i\in\left\{4,8\right\},\,k=9,\\
0 & \text{otherwise}.\end{array}\right.
\end{align*}
One might think about $X^1,\ldots,X^8$
as about price process of $8$ stocks,
while the CIR process
$X^9$ is their common stochastic volatility.
Notice that Feller's condition for $X^9$
is not satisfied ($\frac{2\lambda\bar{v}}{\eta^2}=0.8<1$),
that is, $0$ is accessible boundary point for $X^9$
(with reflecting boundary behaviour).
The discretised process $(X_{\Delta,j\Delta}^9)_{j=0,\ldots,J}$
can become negative.
We, therefore, use the following discretisation scheme
\begin{align*}
X^i_{\Delta,j\Delta}&=X^i_{\Delta,(j-1)\Delta}\left(1+r\Delta+\sigma^i\sqrt{\left(X_{\Delta,(j-1)\Delta}^9\right)^{+}}A^i\sqrt{\Delta}\xi_j\right),\\
X^9_{\Delta,j\Delta}&=X^9_{\Delta,(j-1)\Delta}+\lambda\left(\bar{v}-\left(X_{\Delta,(j-1)\Delta}^9\right)^{+}\right)\Delta+\eta\sqrt{\left(X_{\Delta,(j-1)\Delta}^9\right)^{+}}A^9\sqrt{\Delta}\xi_j,
\end{align*}
where 
$i\in\left\{1,\ldots,8\right\}$ and $x^{+}\doteq\max\left\{x,0\right\}$.
Here, we consider of the functional
$f\left(x\right)=\max\left\{\max_{i\in\left\{1,\ldots,8\right\}}x^i-1,0\right\}$
and, as in Section \ref{subsec:gbm_10}, use the simplified control variate $\tilde{\tilde{M}}_{\Delta,T}^{(1)}$
(we again exclude the RRCV approach).
The results for the log-scaled sample are illustrated in Table~\ref{tab_9d_heston}.
We get that the ECDF for the SRCV approach
has a similar form as the one from
Section~\ref{subsec:gbm_10}
(see the sixth plot in Figure~\ref{fig:1}).
Notice that the values of the estimators lie in all cases around 4.6 (SMC: 4.62, RCV: 4.59, SRCV: 4.60). Nevertheless, in the case of the SRCV approach $75.5\%$ of the paths are located within the interval $\left(3,6\right)$, whereas in case of the SMC approach this holds for only $13.0\%$ of the paths and in case of the RCV approach for only $9.9\%$. This is a further indication of a better numerical performance of the SRCV approach. 
\begin{table}[htb!]
\small
\centering
\begin{tabular}{|l|l|l|l|l|l|}
\hline
Approach & \multicolumn{1}{l|}{Min} & \multicolumn{1}{l|}{Max} & \multicolumn{1}{l|}{Variance} & \multicolumn{1}{l|}{Time (sec)} & \multicolumn{1}{l|}{$\theta$} \\ \hline
SRCV & -6.4 & 2.6 & 50.1 & 444.7 & 0.09 \\ \hline
RCV & -10.2 & 1.0 & 3208.8 & 328.6 & 4.33 \\ \hline
SMC & -1.7 & 9.8 & 1478.8 & 164.5 & 1 \\ \hline
\end{tabular}
\caption{Results of the algorithms for a Call-on-max-option in a high-dimensional Heston model.}
\label{tab_9d_heston}
\end{table}

\section{Proofs}
\label{sec:proofs}
\subsection{Proof of Lemma~\protect\ref{th:2104a1}}
Applying Theorem~11.3 in~\cite{gyorfi2002distribution} and using Assumption~(A1) leads to 
\begin{align*}
&\E\|\tilde h_{j,y,z}-h_{j,y,z}\|^2_{L^2(\PP_{\Delta,j-1})}\\
\le&
 \tilde{c}\, A^2(\log (Np_m(y))+1)\frac{K}{Np_m(y)}+8\inf_{g\in\Psi_{K}}\int_{\mathbb{R}^d}\left(h_{j,y,z}\left(x
\right)-g\left(x\right)\right)^2\,\PP_{\Delta,j-1}(dx),
\end{align*}
since the expected number of (training) paths given $\xi_j=y,V_j=z$ is $Np_m(y)$ and
\begin{align*}
\Var[q_j(X_{\Delta,j\Delta})|X_{\Delta,(j-1)\Delta}=x,\xi_j=y,V_j=z]=\Var[q_j(\Phi_\Delta(x,y,z))]=0.
\end{align*}
By means of Assumption~(A2) and $\log p_m(y)< 0$ we finally obtain~\eqref{eq:2104a2}.

\subsection{Proof of Lemma~\protect\ref{th:2105a1}}
Let us first recall that
$2^{m(m-1)/2}p_m(y)=\PP(\xi_j=y)
=\prod_{i=1}^m \PP(\xi_j^i=y_i)$
(cf.~\eqref{eq:3112a1}).
Lemma~\ref{th:2104a1} together with
formulas~\eqref{eq:2512b2} and~\eqref{eq:2104a2}
as well as $(\sum_{i=1}^{c_m} b_i)^2\le c_m\sum_{i=1}^{c_m} b_i^2$ yield
\begin{align*}
&\E\|\tilde a_{j,o,r}-a_{j,o,r}\|^2_{L^2(\PP_{\Delta,j-1})}\\
&\le c_m\sum_{y\in\{-\sqrt3,0,\sqrt3\}^m}\;
\sum_{z\in\{-1,1\}^\cI}
\left[
p_m(y)
\prod_{i=1}^m H_{o_i}(y_i)
\prod_{(k,l)\in\cI} z_{kl}^{r_{kl}}
\right]^2 \E\|\tilde h_{j,y,z}-h_{j,y,z}\|^2_{L^2(\PP_{\Delta,j-1})}\\
&\le c_m2^\frac{m(m-1)}{2}\sum_{y\in\{-\sqrt3,0,\sqrt3\}^m}\;
\left[
\prod_{i=1}^m H_{o_i}(y_i)
\right]^2\left(\tilde{c}\, A^2(\log N+1)\frac{Kp_m(y)}{N}
+\frac{8\,C_\kappa p_m(y)^2}{K^\kappa}\right)\\
&=
c_m\tilde{c}\, A^2(\log N+1)\frac{K}{N}\E\left[
\prod_{i=1}^m H_{o_i}(\xi_j^i)
\right]^2
+\frac{8\,C_\kappa }{K^\kappa}\!\!\sum_{y\in\{-\sqrt3,0,\sqrt3\}^m}\;
c_m2^\frac{m(m-1)}{2}\left[
p_m(y)\prod_{i=1}^m H_{o_i}(y_i)
\right]^2\\
&=c_m\tilde{c}\, A^2(\log N+1)\frac{K}{N}
+\frac{8\,C_\kappa}{K^\kappa}C_{m,o},
\end{align*}
where in the last equality we used that
$\xi_j^1,\ldots,\xi_j^m$
are independent and all $H_{o_i}(\xi_j^i)$
have unit $L^2$-norm.

\subsection{Proof of Theorem~\protect\ref{th:2106a1}}
It holds
\begin{align*}
\Var[f(X_{\Delta,T})-\tilde M_{\Delta,T}^{(2)}]
&=\Var[M_{\Delta,T}^{(2)}-\tilde M_{\Delta,T}^{(2)}]\\
&=\EE\Var[M_{\Delta,T}^{(2)}-\tilde M_{\Delta,T}^{(2)}|D_N^{tr}]
+\Var\EE[M_{\Delta,T}^{(2)}-\tilde M_{\Delta,T}^{(2)}|D_N^{tr}].
\end{align*}
Due to the martingale transform structure
in~\eqref{eq:2512b1} and~\eqref{eq:0101a1},
we have
\begin{align*}
\EE[M_{\Delta,T}^{(2)}-\tilde M_{\Delta,T}^{(2)}|D_N^{tr}]=0.
\end{align*}
Together with the fact that the system
$
\left\{
\prod_{i=1}^m H_{o_i}(\xi_j^i)
\prod_{(k,l)\in\cI}(V_j^{kl})^{r_{kl}}:
(o,r)\in\cU
\right\}
$
is orthonormal in $L^2$, we get
\begin{align}\label{eq:0101a2}
\Var[f(X_{\Delta,T})-\tilde M_{\Delta,T}^{(2)}]=
\sum_{j=1}^J
\sum_{(o,r)\in\cU}
\E\|\tilde a_{j,o,r}-a_{j,o,r}\|^2_{L^2(\PP_{\Delta,j-1})}.
\end{align}
With the expression $C_{m,o}$
of Lemma~\ref{th:2105a1} we compute
\begin{align*}
\sum_{(o,r)\in\cU}C_{m,o}
&=
\sum_{o\in\{0,1,2\}^m}
\sum_{r\in\{0,1\}^\cI}
C_{m,o}-\sum_{y\in\{-\sqrt3,0,\sqrt3\}^m}\;
c_m 2^\frac{m(m-1)}{2} p_m(y)^2\\
&=
\sum_{o\in\{0,1,2\}^m}
2^{\frac{m(m-1)}2}
C_{m,o}-\sum_{y\in\{-\sqrt3,0,\sqrt3\}^m}\;
c_m 2^\frac{m(m-1)}{2} p_m(y)^2\\
&\doteq\alpha-\beta,
\end{align*}
where $\alpha$ (resp.~$\beta$)
denotes the first (resp.~second)
big sum in the above expression.
Let us compute $\alpha$ and~$\beta$.
Recalling that
\begin{align*}
2^{m(m-1)/2}p_m(y)=\PP(\xi_j=y)
=\prod_{i=1}^m \PP(\xi_j^i=y_i),
\end{align*}
we get
\begin{align*}
\alpha
&=
c_m
\sum_{o\in\{0,1,2\}^m}
\sum_{y\in\{-\sqrt3,0,\sqrt3\}^m}\;
\prod_{i=1}^m
\left[
\PP(\xi_j^i=y_i) H_{o_i}(y_i)
\right]^2
\\
&=
c_m
\left(
\sum_{o_1\in\{0,1,2\}}
\sum_{y_1\in\{-\sqrt3,0,\sqrt3\}}
\left[
\PP(\xi_j^1=y_1) H_{o_1}(y_1)
\right]^2 \right)^m
=c_m,
\end{align*}
where the last equality follows
by a direct calculation.
Recalling that $c_m=3^m 2^{m(m-1)/2}$
(we consider second order schemes),
we obtain
\begin{align*}
\beta
=3^m\!\!
\sum_{y\in\{-\sqrt3,0,\sqrt3\}^m}\;
\prod_{i=1}^m
\PP(\xi_j^i=y_i)^2
=3^m
\left(
\sum_{y_1\in\{-\sqrt3,0,\sqrt3\}}
\PP(\xi_j^1=y_1)^2
\right)^m
=\left(\frac32\right)^m.
\end{align*}
Thus,
\begin{align*}
\sum_{(o,r)\in\cU}C_{m,o}=c_m-\left(\frac32\right)^m=\tilde c_m.
\end{align*}
The last expression
together with Lemma~\ref{th:2105a1}
and~\eqref{eq:0101a2}
yield the result.

\bibliographystyle{model2-names}
\bibliography{ml_vr_bibliography}

\end{document}